# *Surveying 5G Techno-Economic Research to Inform the Evaluation of 6G Wireless Technologies*

Edward J. Oughton, *Member, IEEE*, and William Lehr, *Senior Member, IEEE*

*Abstract*—Techno-economic assessment is a fundamental technique engineers use for evaluating new communications technologies. However, despite the techno-economics of the fifth cellular generation (5G) being an active research area, it is surprising there are few comprehensive evaluations of this growing literature. With mobile network operators deploying 5G across their networks, it is therefore an opportune time to appraise current accomplishments and review the state-of-the-art. Such insight can inform the flurry of 6G research papers currently underway and help engineers in their mission to provide affordable high-capacity, low-latency broadband connectivity, globally. The survey discusses emerging trends from the 5G techno-economic literature and makes five key recommendations for the design and standardization of Next Generation 6G wireless technologies.

*Index Terms*—5G, 6G, Cellular, Mobile, Techno-Economics, Standardization, Wireless.

## I. INTRODUCTION

A new generation of cellular technology has arrived each decade since the first (1G) networks were commercially deployed in the 1980s. We are now in the early stages of the commercial launch of the fifth generation (5G) of cellular technology, and at the start of the research and development lifecycle for the sixth generation (6G) of cellular technologies.

As of 2022, Mobile Network Operators (MNOs) around the globe have deployed Non-Standalone 5G services, and although coverage and adoption numbers are growing rapidly, adoption in many markets still remains modest [1]. Currently none of the commercially available offerings deliver the complete range of 5G performance metrics and use cases targeted by the ITU's IMT2020 vision [2], with all 5G capabilities unlikely to emerge until after later 3GPP releases come to full fruition in the mid-2020s [3].

Unfortunately, there is still uncertainty around the 5G investment proposition for mobile operators, in terms of how they can viably deploy new cellular infrastructure beyond high traffic urban areas. This contrasts with previous generations, such as 4G, where there was a much clearer investment case. For example, the sale of a 4G cellular connection combined with a smartphone enabled users to access mobile web browsing (the 'killer app'), with Apple's iPhone being the key driver of this trend [4]. However, such a clear case is still yet to emerge for 5G. Yet, industry participants are already designing usage cases, performance requirements, and technical specifications to support the next generation ('Next-G') of 6G wireless networks [5]–[8].

This is a familiar, if messy, process that has been repeated as successive lifecycles of wireless technologies migrate from development labs through commercial deployment to eventually become legacy technologies [9]–[12]. Although we are not close to the end of 5G's lifecycle, we are at the end of 5G's beginning, which presents us with an opportune time to mine the lessons learned as we start to formulate plans for the next generation of wireless technologies, many of which will become candidate 6G technologies.

Techno-Economic Assessment (TEA) is the economic evaluation of an engineered system. TEA is important because once fundamental engineering research has been standardized into a new cellular generation, market forces govern the subsequent design, deployment, and success of these wireless technologies [13]. Indeed, strategic telecommunications decisions depend on using rigorous and robust techno-economic analysis to inform network architecture design, network operations, and more broadly the chosen business model and level of investment [14] to deliver Next-G communications services. Understandably, most 5G research has focused on the technical engineering and computer science aspects, but to the potential detriment of not having as sophisticated understanding of the cost implications of different technical developments [15], which are still being established.

Equivalently, much of the non-technical research on wireless networks has been inadequately informed by the technical details of emerging technologies. For example, in 5G this paradigm of new communications technologies is driven by increased virtualization, based on the 'cloudification' of both Radio Access Networks (RAN) and other transport network segments (fronthaul, backhaul, core etc.). Yet, this is highly technical detail for business analysts to grasp, increasing the likelihood of poor decision making (thereby, prioritizing the need for engineers to have a good grasp of techno-economics).

'Techno-economics' research must be well grounded both in the technical engineering details of the wireless technologies being evaluated, and also in the market economic considerations that reflect how much new features cost relative to the service enhancements they may provide to consumers and

This work was supported by the International Monetary Fund's Digital Infrastructure Costing Estimator (DICE) project and the U.S. National Science Foundation under Grant 2037777.

Edward J. Oughton is with the College of Science at George Mason University, Fairfax, VA 22030 USA, and the University of Oxford, Oxford, UK (e-mail: eoughton@gmu.edu).

William Lehr is with the Computer Science and Artificial Intelligence Laboratory at the Massachusetts Institute of Technology, Cambridge, MA 02139 USA (e-mail: wlehr@mit.edu).

businesses [16]. Achieving this can help bridge the gap between the engineering-aspects of wireless technology (such as in 5G or 6G) and the market deployment conditions in which they will be operated.

TABLE I
LIST OF ACRONYMS

| Acronym | Full Term |
|---|---|
| 3GPP | 3rd Generation Partnership Project |
| 5G | Fifth Generation |
| 5G NR | Fifth Generation New Radio |
| 6G | Sixth Generation |
| ARPU | Average Revenue Per User |
| Capex | Capital Expenditure |
| COTS | Commercial Off The Shelf |
| C-RAN | Cloud Radio Access Network |
| DCA | Discounted Cash Analysis |
| eMBB | Enhanced Mobile Broadband |
| EPC | Evolved Packet Core |
| FTTP | Fiber To The Premises |
| GUI | Graphical User Interface |
| HetNet | Heterogenous Network |
| IEEE | Institute of Electrical and Electronics Engineers |
| IIoT | Industrial Internet or Things |
| IoT | Internet of Things |
| IRR | Internal Rate of Return |
| LMIC | Low and Middle Income Country |
| LTE | Long Term Evolution |
| LTE-A | Long Term Evolution Advanced |
| MIMO | Multiple In Multiple Out |
| MIoT | Massive Internet of Things |
| mMIMO | Massive Multiple In Multiple Out |
| mMTC | Massive Machine Type Communication |
| mmW | Millimeter wave |
| MNO | Mobile Network Operators |
| NB-IoT | Narrow Band Internet of Things |
| Next-G | Next Generation |
| NFV | Network Function Virtualization |
| NPV | Net Present Value |
| Opex | Operational Expenditure |
| PoP | Point of Presence |
| RAN | Radio Access Network |
| ROI | Return On Investment |
| SDN | Software Defined Networking |
| TCO | Total Cost of Ownership |
| TEA | Techno Economic Analysis |
| TWDM-PONs | Time and Wavelength Division Multiplexed Passive Optical Networks |
| UAV | Unmanned Aerial Vehicle |
| UE | User Equipment |
| uRLLC | Ultra Reliable Low Latency Communication |

The goal of this paper is to provide a literature survey of the techno-economics research over the past decade that contributed to the development, standardization, and deployment of 5G. While this is a highly useful endeavor in isolation, we synthesize the findings into a set of recommendations for the future of TEA, particularly regarding Next-G wireless technology development. Specifically, the research questions this survey will investigate include:

1. What are the major trends in the use of techno-economic methods for assessing 5G?
2. What worked and what did not, in the use of these methods for assessing the techno-economics of 5G?
3. How can the use of techno-economic methods be improved for evaluating candidate 6G technologies?

Before embarking on this survey, we first provide a quick context-setting review of the status of 5G technologies and use cases in Section II. We then define what is meant by 'techno-economic' research, including outlining the standard steps of TEA (Section III). The main meta-review of the 5G techno-economic literature is undertaken in Section IV, with Section V summarizing the characteristics of the sample of papers evaluated. Finally, in Section VI the research findings are discussed, and recommendations made for how these research methods may be improved in the future. Conclusions are given in Section VII.

II. THE 5G CONTEXT, USE CASES AND KEY TECHNOLOGIES

Throughout the first four generations of cellular technologies, culminating in the 4G era (2010 onwards), the principal business model adopted by MNOs focused on selling consumer subscriptions based on the quantity of monthly traffic usage per subscriber [17]. However, with the number of human subscribers maximized in most markets and a shift towards unlimited data packages, operators have been shifting their focus to identify new revenue streams, potentially via new 5G services.

To achieve these revenue gains, MNOs are hoping to target 'verticals', which refer to the use of 5G services across a range of industrial sectors [18]. Relative to previous generations, this is seen as a potentially disruptive change in terms of how businesses adopt and use wireless connectivity [19]. Indeed, this is a shift from the previous focus of operators on providing consumer broadband services. The vertical sectors being targeted include energy [20], transport and logistics [21], [22], healthcare [23], [24], live events [25], manufacturing [26], [27], agriculture [28], construction [29], and tourism [30]. Yet, deployment of these new wireless 5G services require bespoke forms of software, hardware and spectrum to meet the connectivity requirements of every vertical sector business model [31]–[33].

The success of this ambition for each industrial sector will depend on its technical and economic viability. This viability reflects the cost of delivery versus the perceived value that businesses and consumers are Willing-To-Pay [34]. Without a clear business case both investors and network operators will not undertake the necessary investment to deploy new 5G services [35]. This highlights the important need for high-quality TEA.

The key technical characteristics of 5G have emerged over the past decade to support three main use cases, identified as Enhanced Mobile Broadband (eMBB), Massive Machine Type Communication (mMTC), and Ultra Reliable Low Latency Communication (uRLLC). Of these three, only the first has been demonstrated to be broadly successful for MNOs. The market potential for mMTC and uRLLC and the role that MNOs will play in those markets should they develop as hoped remains speculative.

*Enhanced Mobile Broadband* is an extension of the highly successful Mobile Broadband use case in 4G, which in combination with the rise of smartphones, propelled mass media content, applications, and mobile services into the palms of billions of smartphone and tablet users worldwide. However, in 4G the rapid adoption of smartphones and use of mobile broadband services meant cellular networks were frequently saturated by demand in high user density hotspots, particularly in busy periods of the day (e.g., 5 pm). The aim of eMBB is to dramatically increase the data rate per cell over a wider coverage area by increasing spectral efficiency, in order to support many more users simultaneously demanding high-quality multimedia content [36], [37]. Delivering eMBB is seen as the first phase in deploying 5G infrastructure and services, as it is possible to use Non-Standalone 5G, with the 5G New Radio (5G NR) interface implemented, while reusing the 4G Evolved Packet Core (EPC). Thus, the 5G business case for eMBB by MNOs is well-demonstrated both by the need to reduce capacity provisioning-costs and to keep up with competitive pressures and consumer demand for ever-more data.

*Massive Machine Type Communication* (mMTC) is synonymous with the much widely known concept of the Internet of Things (IoT) [38], but with particular relevance in 5G towards industrial use by 'verticals', also known as the Industrial Internet or Things (IIoT) [39]–[41]. Many industries previously may have relied on predominantly fixed or Wi-Fi connectivity to utilize sensors throughout the production process, but new developments have taken place in this domain [42]. Particularly the availability of local private spectrum licenses has increased the attractiveness of cellular connectivity, especially those business models where mobility within a production facility is essential, and existing communications options may not perform as desired when using unlicensed spectrum bands [43], [44]. Despite the significant progress made in this area, the business case for wireless IoT applications, the role of MNOs and cellular 5G technologies in meeting this market opportunity is still developing. Markets for IoT devices and supporting application software and hardware integration remain fragmented, and their value can be highly dependent on the business model of the potential use case.

*Ultra-Reliable Low Latency Communication* (uRLLC) is featured in 5G NR to provide wireless connectivity for those use cases which have very stringent reliability and latency requirements [45], [46]. These use cases could include types of industrial automation or transport systems [47] where transfer of information may be time sensitive (e.g. 1 ms latency) with high reliability (i.e., a low failure rate) [48], [49], but also requiring high-speed mobility, such as Unmanned Aerial Vehicles (UAVs) [50]–[52]. This is arguably one of the most novel features of 5G when compared to the systems that came in generations before this technology [53], but wide-spread commercial realization of networks that can meet these stringent performance requirements are likely to depend on future 3GPP standard releases (e.g. implementation of Release 16); and will contend with alternative (non-cellular) technical options. Of the three focal market opportunities targeted by 5G usage cases [2], uRLLC is the one that is furthest removed from commercial realization so far.

The technical design of 5G for all three of these usage cases has benefited from, and depends on, a wide range of R&D technical innovation progress made over the past decade. Given there are three main ways to expand the capacity of a wireless network, these innovations are presented regarding these options.

Firstly, capacity can be greatly enhanced through increasing the density of cells and enabling greater spectrum reuse. A good example is the growth of small cell deployment [54], [55] which is used to significantly enhance the available capacity compared to a traditional approach using only macro cells [56], [57]. Indeed, the use of Heterogenous Networks (HetNets) in 5G [58]–[60] tie into the increased use of virtualization via Software Defined Networking and Network Function Virtualization (SDN/NFV), where such capabilities can enable network slicing to vertical sectors with different quality of service requirements [61]–[64]. Often this is referred to by the term 'flexibility', which reflects the providing of new capacity via state-of-the-art 'softwarized' networks [65].

Secondly, one can add supplementary spectrum bandwidth to simultaneously provide more frequencies for packet transfer. Indeed, there has been considerable experimental research examining the use of millimeter wave spectrum, which will inform the use of very large bandwidths of spectrum above 26 GHz [66], [67]. Already, the deployment of midband spectrum has become central to 5G deployments in markets across the world, particularly in the 3.5 GHz band [68], demonstrating the importance of this approach to improving capacity.

Finally, one can provide greater spectral efficiency. Since 4G LTE, the ability to fit more packets on a radio wave has been slowing, with research indicating [69] the Shannon Bound has effectively been reached. However, thankfully in 5G the deployment of much higher order Massive Multiple-In, Multiple-Out (mMIMO) radio technology [70] provides the possibility to greatly improve wide-area network spectral efficiency [71], [72]. Indeed, this is one of the main tools in the fifth generation of cellular technologies [73].

Before turning to our review of the literature, let us explain and define what we mean by 'techno-economics' research.

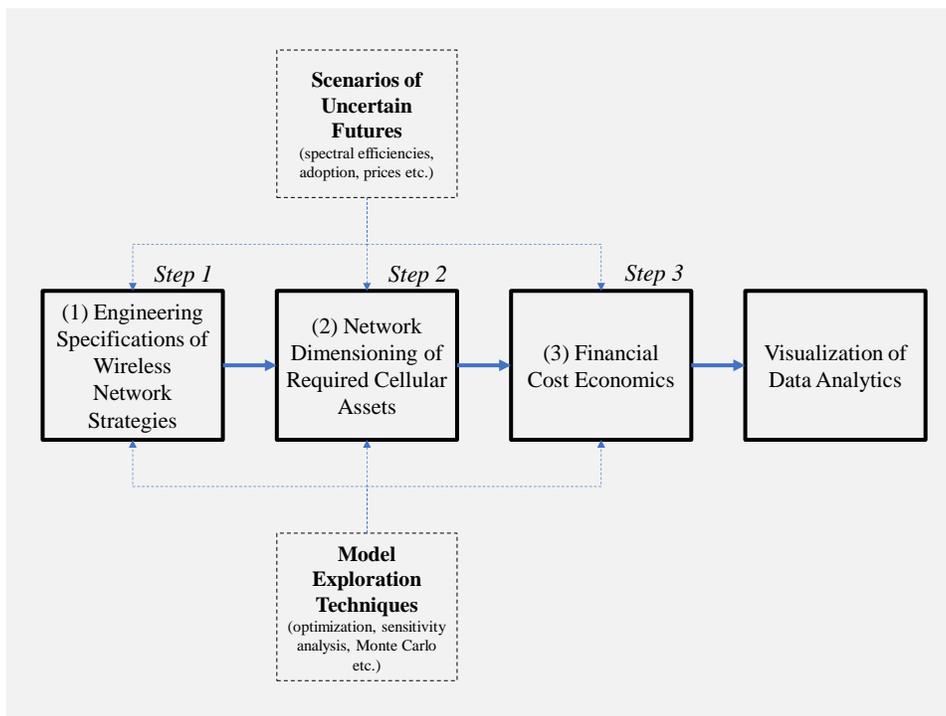

**Fig. 1.** Defining the process of TEA.

### III. DEFINING TECHNO-ECONOMICS

Herein, we restrict our consideration of 'techno-economics' to the quantitative modeling approach adopted in engineering disciplines and variously known as 'techno-economic assessment,' 'techno-economic analysis,' (TEA), or 'techno-economic modeling' (TEM) [74]. The aim of such an assessment paradigm is to be able to quantitatively evaluate the economic performance of different types of engineered systems, as well as quantify other impacts, for example, relating to energy consumption or environmental emissions. TEA techniques have been widely applied to model new technologies, including for 3G [75], 4G LTE [76]–[78], Wi-Fi [79], satellite broadband [80]–[83], WiMAX [84], femtocells [85], [86], point-to-point wireless backhaul [87], Intelligent Transport Systems [88], industrial networks [89], spectrum sharing [90], [91], smart meters [92], and fixed broadband [93], [94].

We constrain our analysis here to focus on TEA papers which solely appraise 5G. Fig. 1 illustrates a general theoretical overview for how TEA is applied within engineering, specifically regarding the assessment of cellular systems, such as 5G and (in the future) 6G.

Traditionally the area of TEA commonly made use of spreadsheet modeling, but increasingly software- based approaches are used to provide greater flexibility and rigor in the modeling process, particularly regarding the use of advanced techniques such as sensitivity analysis or Monte Carlo simulation. The 'economic' aspect is developed in terms of modeling both the costs of supply and the potential demand for the goods and services being evaluated via a specific engineered system. Supply-side costs are frequently captured in terms of the mix of capital and operational costs involved with production, and how they combine to provide insight into the Total Cost of Ownership (TCO) over the lifetime of a particular set of assets. Often the TCO is obtained by carrying out a Discounted Cash Analysis (DCA) to reduce a set of investments over a time-period to a specific Net Present Value (NPV). Such costs can then be related to the potential demand-side revenue which can be obtained, providing insight on the Return on Investment (RoI) and Internal Rate of Return (IRR).

Three main steps are identified based on existing theory. First, in *Step 1* a set of engineering designs are specified which represent different cellular technologies. Many assessments focus on modeling the capacity and coverage of a particular cellular system for different quality of service levels. For example, capacity-focused engineering evaluations focus on (i) the spectral efficiency of a cellular technology, (ii) the available spectrum bandwidth for a carrier channel being modeled, and (iii) the cellular density of existing sites and their level of sectorization. Some studies also include energy consumption and may focus on quantifying the level of electricity consumption of specific technologies.

Second, in *Step 2* the number of network assets and any other affiliated components are quantified, for a particular quality of service achieved for the engineering specifications defined in Step 1. Often this involves specifying both a 'bill of works' and a 'bill of materials', with these quantities usually linked into a specific statistical framework reflecting spatial and temporal reliability criteria.

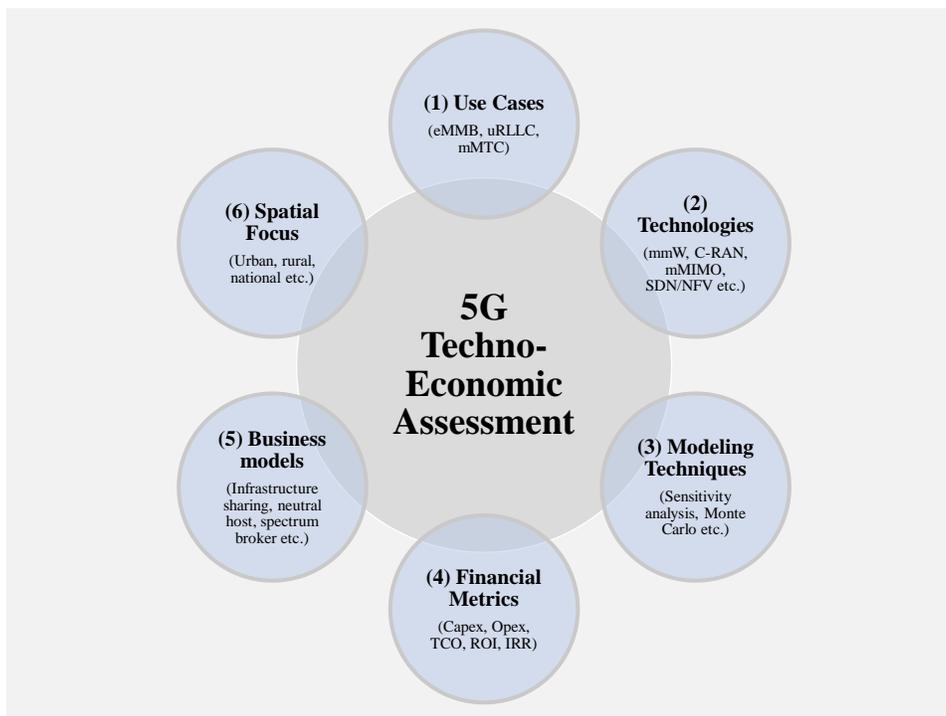

**Fig. 2.** Identifying the six main topics for 5G TEA.

Finally, in *Step 3* a set of corporate finance techniques are usually carried out to assess the range of capital and operational expenditures involved with building a particular wireless network architecture. The TCO is ideally used to capture the range of Capex and Opex payments incurred over the asset and network lifetime. Often this includes producing metrics which relate to demand, such as the RoI.

The TEA models are of special relevance to stakeholders interested in assessing the commercial viability of investing in and deploying services or offering products making use of the modeled technology. The results of such models provide inputs to strategic plans and investment decisions, as well as assessments of competition and market dynamics.

Before moving on to our review of the 5G TEA literature, it is worth mentioning that other quantitative methods can be regularly employed by stakeholders to inform economic and policy decision-making related to Next-G wireless technologies, which are out of the scope of this current survey. These methods include econometrics, systems dynamics, and various competitive strategy frameworks [95], [96].

IV. REVIEWING 5G TECHNO-ECONOMIC ASSESSMENTS

To conduct a review of the 5G TEA research, we start by selecting a total of 150 potential techno-economic studies and select a sample of 75 published or presented at reputable journals and conferences (Table II and III). To collect the total papers, search terms for relevant terminology (such as 'techno-economics', 'cost' etc.) were applied in English with the term '5G' to both IEEE Xplore and Google Scholar. The results of those searches were manually mined for important 5G techno-economic papers (with attention paid to the authors, citation links, and research focus). We include relevant papers identified by our first-order review of the search results, as well as those second-order papers identified by reviewing manuscripts which referenced the most highly cited 5G TEA papers (via the 'cited by' functionality in Google Scholar).

Papers were excluded if they (i) did not explicitly focus on 5G or key 5G technologies, (ii) did not meet high quality academic standards, (iii) only spoke qualitatively about 5G TEA, (iv) were not clearly peer-reviewed by a journal or conference committee, and/or (v) if they were pre- PhD outputs (e.g., Master's theses were excluded). Where the review identified a paper first presented at a conference, and then later formally published in a journal, we only included the latter peer-reviewed publication, to avoid duplication.

Our analysis of the selected studies focuses on six key topics which cover the main aspects of 5G TEA identified for this analysis, as visualized in Fig. 2. We address each of these in the following sub-sections. Evaluation of these paper is also considered within the broader open science agenda and the growing need for engineers to produce reproducible research [97]–[99], particularly in 5G research focusing on data-driven analysis [100], [101].

*A. Use Cases*

Often the delivery of eMBB is compared using techno-economic methods against existing 4G LTE and LTE-Advanced options [102]. Analyses frequently focus on evaluating the cost efficiency of using legacy technologies against state-of-the-art 5G technologies, such as via millimeter wave [103] or ultra-dense HetNets with small cells [104], either indoors [105], [106] or outdoors [107], [108]. Many of these papers have emerged in the peer-reviewed literature after the standardization of 5G in Release 15, compared to earlier work which occurred before standardization, circa-2014-2018 [109]–[111]. Certainly those papers published later are arguably more detailed and rigorous, given there is more clarity on which technologies would be included in 5G, making it easier to assess the potential techno-economic implications and the types of business cases eMBB may pose for network operators [35].

Assessment has begun for use cases which target Ultra Reliable Low Latency Communications (uRLLC), one of the key features of 5G, which has hitherto received far less attention than eMBB. However, providing the latency targeted by 5G use cases is still a major challenge, particularly in trying to do this in a cost-effective way for mission critical environments which are delay sensitive [112]. What is essential therefore is the ability to cost-effectively optimize the deployment of infrastructure to minimize any potential delays in service without making a network architecture so over-engineered that any configuration fails to be economically viable [113].

One hope is that by utilizing Commercial Off The Shelf (COTS) processing hardware, the costs of delivery can be reduced compared to the use of traditional proprietary technologies from the major equipment vendors [114]. The present challenge moving forward is to be able to develop models which can optimize the network latency for different traffic loads, given the spatial and temporal patterns exhibited throughout the day from heterogeneous User Equipment (UE) patterns [115]. In the short to medium term, low latency communication applications may be more frequently deployed in private industrial networks, which either take place indoor within factories [116], [117], or outdoor in a relatively small industrial area (<10 km$^2$) [118], [119].

There is still debate in the literature as to whether the main way to monetize Next-G cellular generations (e.g. 5G and 6G) will be via human-directed services [120] or via machine-directed services [121], [122]. Often the Massive Machine Type Communication (mMTC) use case is also referred to as the 'Massive Internet of Things' (MIoT). However, each term refers to the need for wide-area coverage of cellular services to support machine usage, beyond the more basic Narrow Band IoT (NB-IoT) standard supported in 3GPP Release 13 [123] based on 4G. Ultimately, these machines could be stationary sensors such as smart meters [92] or devices placed on other moving objects (or even living animals), such as for agricultural applications [124]. Techno-economic use case analysis suggests that designing dedicated cellular IoT networks purely for a single use case is not yet economically viable for many applications due to poor cost-efficiency, supporting the need for traffic aggregation across multiple different uses in vertical sectors [125].

Of the three main use cases for 5G, much of the techno-economic assessment literature in the sample evaluated here focuses exclusively on eMBB (33%), with only a small number evaluating uRLLC (4%) and none focusing entirely on mMTC. This could be driven by the fact that 3GPP standardization has mainly focused on eMBB until recently. Enabling the capabilities of uRLLC and mMTC depends on the finalization of future 3GPP standards (particularly uRLLC features in Release 16). These use cases have arguably been left for later standardization because there needs to be additional innovation and development to enable better coordinated system topologies, ranging from multiple radio access technology for reliable access to intelligent traffic routing [126], to support the quality of service required. In total, only 8% of the studies state they focus on all use cases. Fig. 3 (A) visualizes the quantity of publications published by use case category over time for the selected sample, with the number of outputs annually growing year-on-year, from approximately four per year before 2016, to over ten per year from 2017 onwards.

*B. Technologies*

This paper has already reviewed and identified the main technologies central to 5G, including network densification, millimeter wave, mMIMO and the virtualization paradigm of SDN/NFV cloud processing. We now use these categories to discuss the pertinent 5G TEA literature. Tables II and III provide a summary of the technologies evaluated in the 5G TEA sample.

Firstly, the aim of network densification is to massively increase the system capacity thanks to more spectrum reuse, often by utilizing multi-tier networks, where macro cell sites focus on wide-area coverage and smaller cells provide hotspot capacity in areas of very high traffic demand [127]. For example, techno-economic evaluation of 5G network densification is explored for The Netherlands by quantifying both the potential required demand and consequential costs of supply-side strategies in meeting future traffic loads [128]. Different densification options for the Netherlands focus on deploying a small cell layer either on its own, or by deploying a combination of small and macro cells in a HetNet configuration. In contrast, TEA has also been explored for only macro cell densification by using detailed modeling for eight Low and Middle-Income Countries (LMICs) focusing on comparing 5G strategies to existing 4G options [16]. Given the increased number of RAN assets, such an architecture naturally raises questions about backhaul transport options, with other techno-economic analyses focusing on network cost minimization of fixed optical [129], wireless backhaul [130], and millimeter wave mesh networks [131].

TABLE II
5G TEA PAPERS INCLUDED IN THE SAMPLE (LEAD AUTHORS A-L)

| Author(s) | Year | TEA Topic | Technologies | Use Case | TEA Metric |
|---|---|---|---|---|---|
| Al-Dunainawi et al. | 2018 | Green network costs | C-RAN | Could support all | TCO |
| Andrews et al. | 2017 | Infrastructure sharing | Business models, Infrastructure sharing | Not stated | Profit |
| Araújo et al. | 2018 | Rural network technologies | 5G, FTTx | Not stated | Capex only |
| Arévalo et al. | 2018 | Optical fronthauling in urban areas | Fronthaul, CPRI, DSP, 5G transport networks | Could support all | Capex only |
| Asgarirad et al. | 2021 | 5G fronthaul using FTTH | C-RAN, fronthaul, 5G transport networks | Could support all | Capex only |
| Basta et al. | 2017 | Core network using SDN/NFV | SDN/NFV, 5G transport networks | Could support all | Asset quantification only |
| Basu et al. | 2021 | Controller Architecture for SDN | SDN/NFV, 5G transport networks | uRLLC | TCO |
| Bondarenko et al. | 2019 | Network planning optimization | 5G NSA | eMBB | Profit |
| Bongard et al. | 2020 | Converged wireless-optical access networks | C-RAN | Could support all | TCO |
| Bouras & Kollia | 2020 | mmW vs midband spectrum | mmW, midband | eMBB | TCO |
| Bouras et al. | 2016 | Virtualization | SDN/NFV | Could support all | TCO |
| Bouras et al. | 2015 | DAS, UDNs | DAS, small cells | eMBB | TCO |
| Bouras et al. | 2018 | MIMO, DAS | mMIMO, DAS | eMBB | TCO |
| Bouras et al. | 2020 | Cognitive radio and SDN | Cognitive Radio, SDN/NFV | eMBB | TCO |
| Bouras et al. | 2017 | DAS, UDNs | DAS, small cells | eMBB | TCO |
| Bugár et al. | 2020 | Dynamic operator selection when sharing spectrum | 5G HetNets | Not stated | Prices |
| Cano et al. | 2019 | Infrastructure + service provider game | Network slicing, multi-tenancy, business models | Not stated | Revenue |
| Cavalcante et al. | 2020 | 5G for rural and remote areas | Business models | eMBB | Profit |
| Chen et al. | 2016 | Network planning optimization | Small cells, backhaul | Could support all | Capex only |
| Chiaraviglio et al. | 2017 | Viability in rural and remote areas | UAVs, Large Cells (LC) | eMBB | Prices |
| Chiha et al. | 2020 | In-flight connectivity | Satellite 5G | eMBB | TCO |
| De Souza | 2021 | Rural network technologies | SDN/NFV, business models | eMBB | TCO |
| Degrande et al. | 2021 | Intelligent Transport Systems | ITS road-side systems | Could support all | TCO |
| Gangopadhyay et al. | 2019 | Resilitent cost-effective transport | Cloud, SDN/NFV | Not stated | Capex only |
| Ge et al. | 2019 | Network planning optimization (backhaul) | Small cells, backhaul | Not stated | Capex and opex |
| Gedel and Nwulu | 2021 | Infrastructure sharing | Business models, Infrastructure sharing | Not stated | TCO, ROI |
| Gedel and Nwulu | 2021 | Low latency DWNA | MEC, SDN/NFV, mMIMO, D2D | uRLLC | TCO |
| Ghoreishi et al. | 2020 | Cloud-Based Caching-as-a-Service | Cloud, SDN/NFV, C-RAN | Could support all | ROI |
| Gomez et al. | 2020 | Market-driven resource allocation | SDN/NFV, network slicing | Not stated | Revenue |
| Haddaji et al. | 2018 | BackHauling-as-a-Service | Network slicing, backhaul, multi-tenancy | Could support all | TCO, ROI |
| Haile et al. | 2020 | Network planning optimization (RAN) | Small cells, UDNs | Could support all | TCO |
| Han et al | 2017 | Profit optimization of sliced networks | Network slicing, SDN/NFV | Could support all | Profit |
| Jha and Saha | 2018 | Deployment scenarios | mMIMO, mmW, 5G HetNets | eMBB | TCO, revenue |
| Khalil et al. | 2017 | Rural network technologies | TVWS | eMBB | Capex and opex |
| Kumar et al. | 2021 | Rural network technologies | Network slicing, SDN/NFV, business models | eMBB/uRLLC/mMTC | TCO |
| Kusuma and Suryanegara | 2019 | Urban deployment | 5G HetNets, mmW, Wi-Fi | eMBB | Capex and opex |
| LiWang et al. | 2019 | Offloading optimization for satellite-IoV nets | Connected vehicles, satellites, cloud | Could support all | Prices |
| Luong et al. | 2017 | Economic/pricing model survey | 5G HetNets, mMIMO, mmW, C-RAN | Could support all | Prices |

TABLE III
5G TEA PAPERS INCLUDED IN THE SAMPLE (LEAD AUTHORS M-Z)

| Author(s) | Year | TEA Topic | Technologies | Use Case | TEA Metric |
|---|---|---|---|---|---|
| Martín et al. | 2019 | Use case assessment + optimization | - | eMBB/uRLLC/mMTC | Other |
| Mesogiti et al. | 2020 | 5G optical transport networks | C-RAN, SDN/NFV, 5G transport networks | eMBB/uRLLC/mMTC | TCO |
| Musumeci et al. | 2016 | Small cells, optical interface requirements | Small cells, C-RAN, 5G transport networks | Could support all | Asset quantification only |
| Nikolikj and Janevski | 2015 | Performance evaluation | mmW, Wi-Fi, 5G HetNets | eMBB | TCO |
| Ouamri et al. | 2020 | Coverage + cost optimization | 5G HetNets | Not stated | TCO |
| Oughton and Frias | 2018 | Capacity, coverage and rollout | Small cells, business models, midband | eMBB | TCO |
| Oughton and Jha | 2021 | Capacity, coverage + spectrum policy | 5G HetNets, midband | eMBB | TCO |
| Oughton and Russell | 2020 | Spatio-temporal assessment of deployment strategies | Small cells, 5G HetNets, midband | eMBB | TCO |
| Oughton et al. | 2018 | Scenario assessment of deployment strategies | Small cells, 5G HetNets, midband | eMBB | TCO |
| Oughton et al. | 2019 | Scenario assessment of deployment strategies | Small cells, 5G HetNets, midband | eMBB | TCO |
| Oughton et al. | 2019 | Network planning | 5G HetNets, mmW, midband | eMBB | TCO |
| Oughton et al. | 2021 | Scenario assessment of policy options | C-RAN, SDN/NFV, 5G HetNets, midband | eMBB/uRLLC/mMTC | TCO |
| Paglierani et al. | 2019 | Immersive content via C-RAN | SDN/NFV, C-RAN, small cells | eMBB | Capex and opex, ROI |
| Pavon-Marino et al. | 2020 | Filterless technologies in optical networks | SDN/NFV | Could support all | Capex only |
| Raza et al. | 2015 | Optical transport energy + equipment costs | DWDM, 5G backhaul, 5G transport networks | Not stated | Other |
| Rendon Schneir et al. | 2021 | A business case for 5G services in an industrial sea port area | SDN/NFV, C-RAN, small cells, 5G HetNets | eMBB/uRLLC/mMTC | TCO, ROI |
| Rendon Schneir et al. | 2019 | Business case for an urban area | SDN/NFV, C-RAN, small cells, 5G HetNets | eMBB | TCO, ROI |
| Rianti et al. | 2020 | Network planning | mmW | Not stated | TCO, ROI |
| Roblot et al. | 2020 | Vertical use cases | C-RAN | Could support all | TCO |
| Santoyo-González et al. | 2018 | Latency-aware network planning optimization | SDN/NFV, cloud | uRLLC | Asset quantification only |
| Sevastianov and Vasilyev | 2018 | Pricing optimization | - | Not stated | Prices |
| Smail and Weijia | 2017 | Technology assessment | mmW | eMBB | TCO, ROI |
| Suryanegara | 2018 | Revenue per machine | - | uRLLC/mMTC | Revenue |
| Teixeira et al. | 2020 | mmW 5G small cells | Small cells, mmW, 5G HetNets | eMBB | TCO |
| Udalcovs et al. | 2018 | Fronthaul assessment | C-RAN, CPRI, fronthaul | Could support all | TCO |
| Udalcovs et al. | 2020 | Fronthaul assessment | C-RAN, fronthaul | Could support all | TCO |
| Vincenzi et al. | 2019 | Revenue maximization | Network slicing | Not stated | Revenue |
| Walia et al. | 2020 | Micro-operators in campus LANs | Network slicing | Could support all | TCO |
| Wang et al. | 2017 | C-RAN | C-RAN | Could support all | TCO |
| Wisley et al. | 2018 | Capacity and costs in dense urban areas | mmW, 5G HetNets, UDNs | eMBB | TCO |
| Yaghoubi et al. | 2018 | Transport networks | 5G HetNets, 5G transport networks | Not stated | TCO |
| Yaghoubi et al. | 2020 | Transport networks | 5G HetNets, 5G transport networks | Not stated | TCO |
| Yan et al. | 2017 | Energy efficiency | C-RAN, X-Haul, 5G transport networks | Could support all | Other |
| Yunas et al. | 2016 | Macro vs femto cells in HetNets | 5G HetNets, UDNs | Not stated | TCO |
| Yunas et al. | 2014 | Indoor deployment assessment | 5G HetNets, small cells | Not stated | TCO |
| Yunas et al. | 2015 | Urban dynamic DAS vs legacy macro cells | UDNs, DAS | Not stated | TCO |
| Zhang et al. | 2021 | Multi-Network Access in 5G | Multi-Network Access, network slicing, business model | Could support all | TCO |

Secondly, there is no doubt that the integration of millimeter wave frequencies (~30-300 GHz) into 5G is a major new feature, but it does mean that the different propagation characteristics of these bands will lead to new architectures and deployment scenarios [132], therefore requiring new techno-economic analysis to understand how physical constraints affect the economics of deployment. Indeed, millimeter wave frequencies have higher propagation losses through objects, as well as greater rain and atmospheric absorption, leading to significantly smaller inter-site distances compared to the sub-6 GHz frequencies used elsewhere for 5G. For example, cell ranges are likely to be below 300 meters [133]. Indeed, the deployment of millimeter wave spectrum is frequently used by small (pico) cells which individually cost little (e.g. $7,900) compared to a full macro cell site which can cost hundreds of thousands of dollars. However, network dimensioning suggests a small cell strategy could require almost 800 units per 1 km$^2$, due to the poor propagation characteristics of these frequencies, making this option extremely expensive overall [134]. Such findings are not isolated to a single study. Relative to mid-band spectrum (e.g. 3.5 GHz), techno-economic analysis shows that millimeter wave is considerably more costly to deploy [135], due to poorer propagation conditions. Other papers in the literature point to the ability to provide massive increases in capacity from millimeter wave spectrum (>100 Gbps/km$^2$) when deployed outdoor. However, to provide a network which can deliver a 100-times capacity increase over 4G LTE, the required investment is 4-5 times larger (e.g. to deliver a guaranteed 100 Mbps per user) [136]. As there are not vast difference in equipment prices between cellular generations, whether 5G is more or less expensive than previous generations very much depends on how the techno-economic problem is set up, and the quality of service level which an operator may wish to viably deploy for users (although this is rarely standardized across different studies in the literature) [137].

Thirdly, mMIMO architecture is a key technology which is already in wide deployment for those countries with active 5G deployment. By delivering much higher spectral efficiency, thanks to the multi-path propagation mMIMO exploits, the cost per bit of information transferred can be significantly lowered (which is much needed given the exponential increase in traffic growth). Energy consumption is also a major cost factor that mMIMO HetNet deployments need to consider [138]. For example, out of the analysis that has been carried out for mMIMO, findings suggest that the number of antennas needs to be kept relatively low (e.g. 16T16R or 32T32R) to ensure the TCO of the technology deployment remains viable, otherwise the operational expenditure from energy consumption can increase to levels that are not economically viable [139]. Such findings have also been identified elsewhere in the techno-economic literature with higher and higher order MIMO leading to an unsustainable level of energy demand (which can also have negative environmental impacts should this energy be coming from environmentally unfriendly fossil fuels) [132].

Finally, the shift to virtualized networks is being driven by potential cost savings, but these can only be achieved by balancing the trade-off between the network load from traffic demand and the necessary data center resources required for processing [140]. Using COTS, rather than vendor-specific equipment, could provide a cost efficiency saving. However, the management of network resources in data centers hosting SDN and NFV controllers will be important to balance quality of service with the costs arising from data processing (e.g. energy demand) [141], highlighting the need for techno-economic analysis of this problem. Indeed, the location of the data processing facilities for virtualization is a challenge for both wide-area networks deployed by network operators [142], as well as those private 5G networks which may commonly be deployed indoor [105], [106]. Virtualized network architectures have the potential to save up to 75% of total cost against traditional types for specific network latency and traffic loads [115]. Latency criteria are highly important however, meaning geographic location of processing facilities starts to play a much more important role than in previous network architectures [113]. Techno-economic assessment suggests that although virtualization could lead to a ROI of 29%, making payback on any investment as small as one year, there are many uncertainties which can derail such outcomes, with downside risks including unknown revenues, traffic and cost growth factors [35].

Fig. 3 (B) visualizes the number of papers focusing on different 5G technologies, with the distribution being highly reflective of the key technology drivers discussed here. For example, 5G HetNets (27%), SDN/NFV virtualization (24%), C-RAN (21%) and small cells (20%) topped the ranking with the greatest number of papers addressing the techno-economic implications of these technologies.

*C. Modeling Techniques*

There are several key modeling techniques used to assess the techno-economic implications of 5G technologies and deployment strategies, many of which focus on uncertainty quantification. Thus, this part of the review builds on a key area of the theoretical techno-economic process defined in Fig. 1. In this review, the three areas of focus will be optimization, sensitivity analysis and scenario analysis.

We define optimization as the ability to identify the best possible solution for a particular constrained problem. The application of this approach is especially favored in the network planning literature, with the aim of optimizing the RAN [143]–[145], backhaul [129], [146] and core network architectures [113], [140]. The consequential challenge is in deploying affordable services to consumers which deliver on the full range of advanced 5G technical specifications in the standard, while recognizing the market-driven reality of MNOs needing to make a profitable return. Thus, the use of optimization for techno-economic models has gained traction in 5G pricing studies [147]. For example, 5G features such as network slicing require a flexible and efficient framework of network organization and resource management in order to optimize profits, but how this is carried out in practice is still being worked out by many network operators and analysts [148].

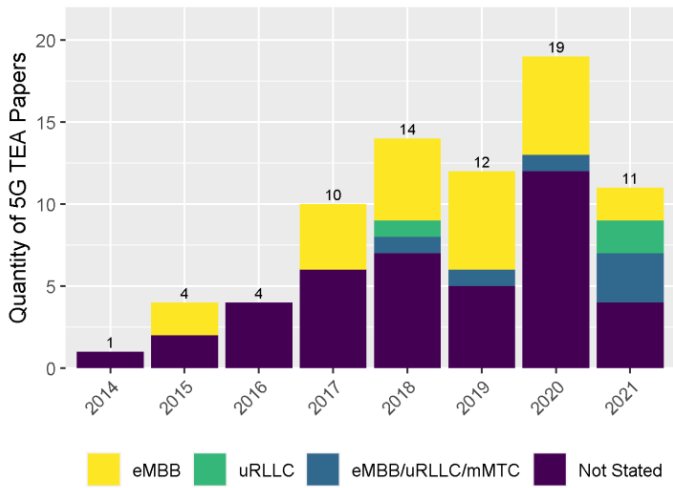
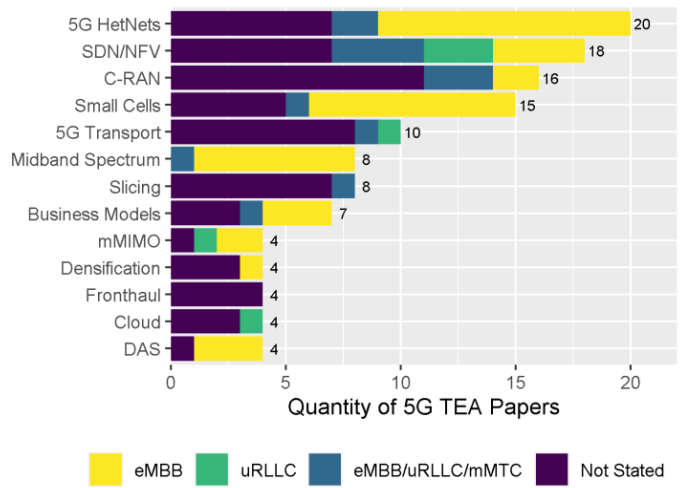
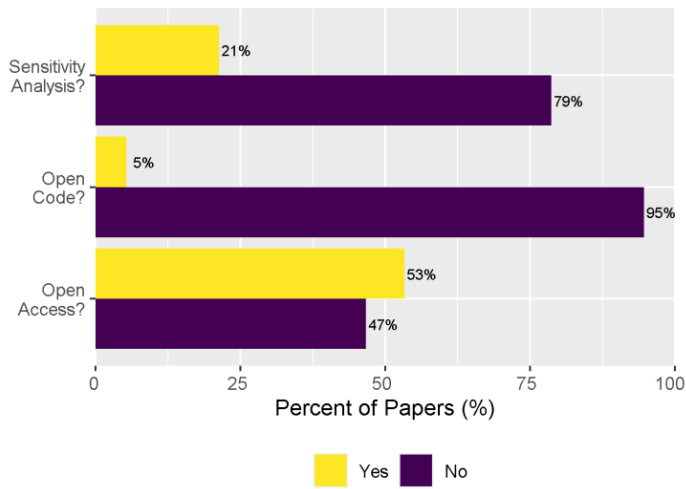
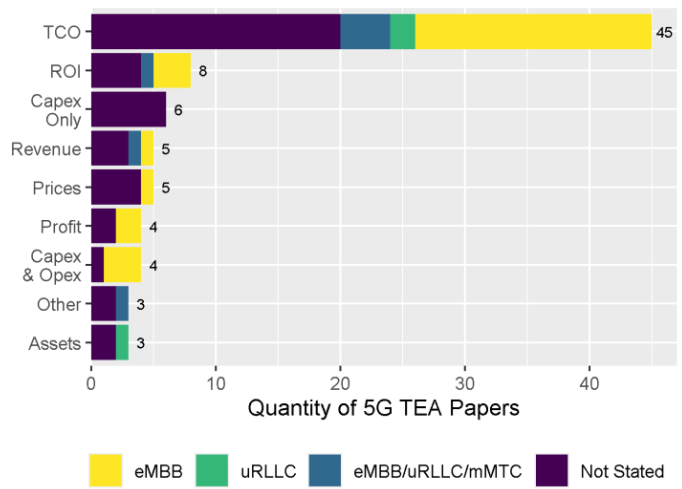
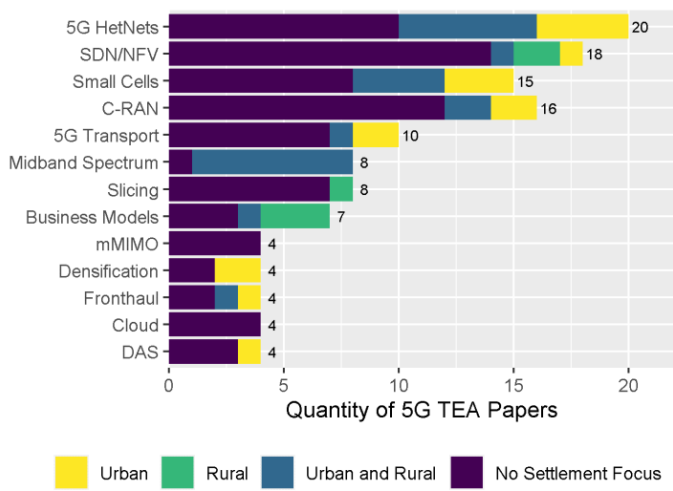
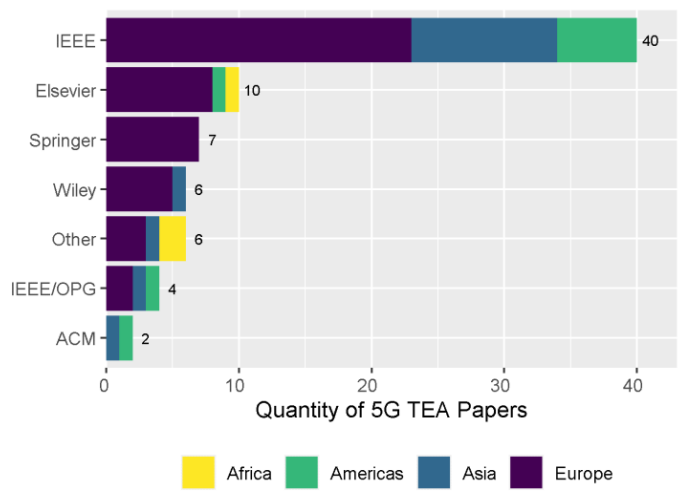

**Fig. 3.** Quantitative summary metrics for the selected 5G TEA sample papers.

Although optimization approaches are common in the techno-economic assessment of communications networks, there is a concern that the structure of the models developed have inherent uncertainties which are often not reported, with analysts either downplaying uncertainty aspects or completely ignoring them. This problem is not necessarily limited to communications engineering, but by failing to address these aspects important information on the model is not reported to decision makers, meaning the techno-economic model insights may be limited, lack robustness or be misleading [149]. To illustrate this point, Fig. 3 (C) indicates that only approximately one fifth of the papers selected in the TEA sample (21%) includes a sensitivity analysis of the developed techno-economic model.

Indeed, we define sensitivity analysis as an uncertainty quantification technique which can be used to explore how specific output variables of interest change as model inputs are repeatedly varied. The process of measuring change in key output metrics can depend on either a limited number of incremental values or ideally by using a full Monte Carlo analysis using continuous distributions. For example, using a techno-economic model of 5G transport networks the sensitivity of the TCO is explored in relation to energy prices, antenna prices, leasing prices and the quantity of available reusable civil infrastructure [130], [150]. It is also common to include sensitivity analysis of the key radio engineering parameters, such as for (i) propagation losses resulting from shadowing and wall penetration [136], or (ii) the ability for 5G networks to cache content to reduce traffic demand [151]. Some studies have included sensitivity analysis of key parameters affecting the deployment of 5G, such as the total spectrum cost [15].

We define scenario analysis as a technique which can be used for strategic planning by subjecting a developed quantitative model to meaningful sets of defined input parameters, to produce analytics on potential futures. Such insight can inform future strategic decisions. Scenarios are used to capture important dimensions such as the future population of potential users, the rate of adoption and future traffic growth [152]. This approach is frequently used when there is not enough scientific information to correctly parameterize a techno-economic model to make robust future predictions. Therefore, scenarios do lend themselves well to usefully representing different futures for network demand [137], [153], as this is ultimately an unknown parameter when network operators are making major decisions about how to deploy a new generation of cellular technology such as 5G. In much of the 3GPP literature the term 'scenario' is used to represent different deployment situations [154], such as 'dense urban' or 'rural', with these terms often driving the techno-economic settlement types researchers use to assess different 5G deployment options [137].

*D. Financial Metrics*

As identified already in Fig. 1, a key part of the techno-economic process is converting the quantity of required engineering assets into different financial cost metrics. Such a step is crucial for providing insight into the economic cost ramifications of different strategic 5G decisions. The degree to which this is undertaken successfully in the literature is mixed, however. For example, one of the best financial cost metrics to use is the TCO which represents the combination of both capital and operational expenditure, over a specific time-period best related to asset lifetime (or in cellular, a generational upgrade which generally happens on a decadal basis). Unfortunately, not all studies actually include Opex, instead focusing on purely Capex [129], [153], [155]–[158]. Not only does such an approach produce only a single part of the picture, which could lead to inefficient decision making, but this problem is exacerbated by the general shift in cost structures in 5G to 'Anything-as-a-Service', such as 'Network-as-a-Service'. For example, one way to reduce up-front costs for network operators is to rent equipment and services from a third-party company, rather than invest in building their own assets to provide desired services. This shift in business model structure has the effect of moving network investments which may traditionally have been made as Capex, if an operator was building their own network, to an Opex payment. In which case, the operator is renting the asset from another service provider. This is common with the shift to virtualized networks where there is no guarantee that network operators will be building their own data processing centers. Instead, operators may prefer to access existing capabilities by major providers who already have competitive advantage in cloud networks such as Google, Amazon, or Microsoft. Fig. 3 (D) illustrates the economic metrics used across the 5G TEA sample, with approximately 8% of papers focusing on Capex-only evaluations.

Although Capex-only quantification provides limited understanding, there are also a range of studies which state they focus on 'cost' by building techno-economic models, but actually stop short of allocating financial numbers to the results [113], [140], [159], [160] (approximately 4% of papers). Granted, required assets numbers still can be used to provide insight, in that more assets will be well correlated with higher required investment, but it leaves researchers having to apply their own sets of costs if they need to use a particular analysis to provide insight on a techno-economic 5G problem.

In addition to using Capex and Opex to estimate TCO, some studies also take cost assessments one step further by using estimates of revenue to generate useful metrics such as RoI (or the IRR) [35], [105], [106], [118], [134], [161], [162], broadly representing 11% of papers. Many in industry and government may favor metrics which capture both the supply-side cost as well as the potential demand-side revenue on the basis that a high TCO does not mean a network strategy is unviable if users have a high Willingness-To-Pay for the service and vice versa. Therefore, being able to show comparative analytics for different 5G network strategies, using investment metrics based on the potential financial return, is highly valuable. A range of papers also focus on estimating prices, revenue, and potential profit, representing 7%, 7% and 5% of the sample, respectively.

*E. Business Models*

A substantial research area has been developing around business models for 5G and the consequential techno-economic implications, particularly around infrastructure sharing (approximately 17% of the reviewed papers). We define business model in the communications context as the way that a network operator configures productive resources to provide communications services to consumers and businesses. The interest in this topic is being driven by the weak economic situation in many telecommunications markets. For example, static and/or declining revenues means there are more challenging economic conditions for network operators for 5G deployment who need to invest large amounts of capital to deliver 5G services.

The 5G TEA literature suggests numerous new business model options which could help reduce costs. Firstly, there has been a substantial number of papers which have focused on the cost reduction benefits of infrastructure sharing, given the different types of sharing that could be possible (passive or active sharing, or geographic sharing such as a shared rural network) [16]. Analysis suggest that infrastructure sharing can be beneficial to a network operator, even when the operator potentially has the power to drive a competing operator out of the market [163]. However, imbalances in the resources available to different network operators can complicate the ability to find outcomes which benefit all market participants [164]. Importantly, assessments of infrastructure sharing are not necessarily confined to high-income markets [35], [108], with assessments also focusing on passive sharing in emerging markets, such as in Ghana [165], with the aim of reducing cost and overcoming coverage disparities for lower-income areas.

With network slicing being a key development in 5G this is affecting the current and future business models that operators are using to sell communications services. The slice properties which an operator develops on its network affects the potential expenditure and revenue, with past research focusing on the ramifications of these technical characteristics [148] (such as whether an operator even chooses to build its own network at all). For example, one hope is that by having rural network operators run a single slice of a network, this may reduce the costs for deployment, making a more scalable and viable business model. It is hoped this innovation exhibits greater financial sustainability over the long term compared to existing approaches [166]. Thus, a situation develops where there are both infrastructure providers and service providers, with neither necessarily providing both infrastructure and services, but specializing in one area [167]. Analysis of the techno-economics of such situations is central to understanding the viability of business model options.

*F. Spatial Focus*

The setting in which a 5G network is being deployed has a profound impact on the network architecture design and thus the techno-economics of infrastructure investment. Many TEA papers explicitly focus on a particular spatial perspective, whether that is deploying 5G in dense urban locations with extremely high traffic loads or examining the implications of 5G deployment in challenging rural and remote locations. Therefore, Fig. 3 (E) illustrates the settlement focus of papers based on the main technologies considered in the 5G TEA sample. Approximately 12% of papers focus on both urban and rural deployment, while 9% focus exclusively on urban and 5% exclusively on rural, with the remaining having no spatial focus.

Assessments focusing on urban areas often consider the cost saving benefits of deploying new assets using a virtualized C-RAN, with central London being one case study example [35]. In dense urban areas this is a logical way to design and operate a 5G network as the high traffic load provides strong economics for fiber deployment, which is required in order to fronthaul traffic from remote radio heads back to a shared processing facility [155]. When using 4G macro cellular sites, the costs of deploying a network to meet increasing traffic demand is the main reason why there is so much interest in 5G deployment in urban areas, as these new technologies hope to lower the cost per bit of information transfer [168]. The improvements in spectral efficiency by 5G NR, as well as other new technologies such as mMIMO, make 5G options well suited to deployment in very high population density cities, as has been explored for three metropolitan areas in Indonesia [103]. Although even with these spectral efficiency improvements over 4G, prudent spending will be necessary by network operators to ensure that substantial capacity and coverage improvements can be made over previous generations, without 5G costs increasing to an unviable level for consumers of cellular communications services [136].

In contrast, there is more limited assessment of rural areas. This is surprising given the quantity of citizens presently unconnected to the Internet residing in rural and remote locations. Indeed, almost 3 billion people globally are currently still not connected to a decent broadband connection, providing a clear motivation for trying to better serve this portion of the global population (which amounts to almost 50%). Granted, many of these people may reside in areas already covered by cellular services, thus the barrier may be the adoption of a smartphone. But a large portion are in 'not spots' which are not currently covered by cellular infrastructure, therefore with no 4G or 5G broadband service.

Often a main challenge is whether 5G will be feasible to deploy in these areas where incomes are very low, meaning modest Average Revenue Per User (ARPU). In this circumstance, network operators fail to build a decent business case for making key investments, given the potentially high cost of building a 5G network, particularly in areas where electricity access can be challenging, and existing fiber optic Point of Presence (PoP) density is low [169]. This provides a strong case for the need for more TEA modeling, as this activity can help to identify suitable cost-efficient strategies for reducing the 'digital divide'.

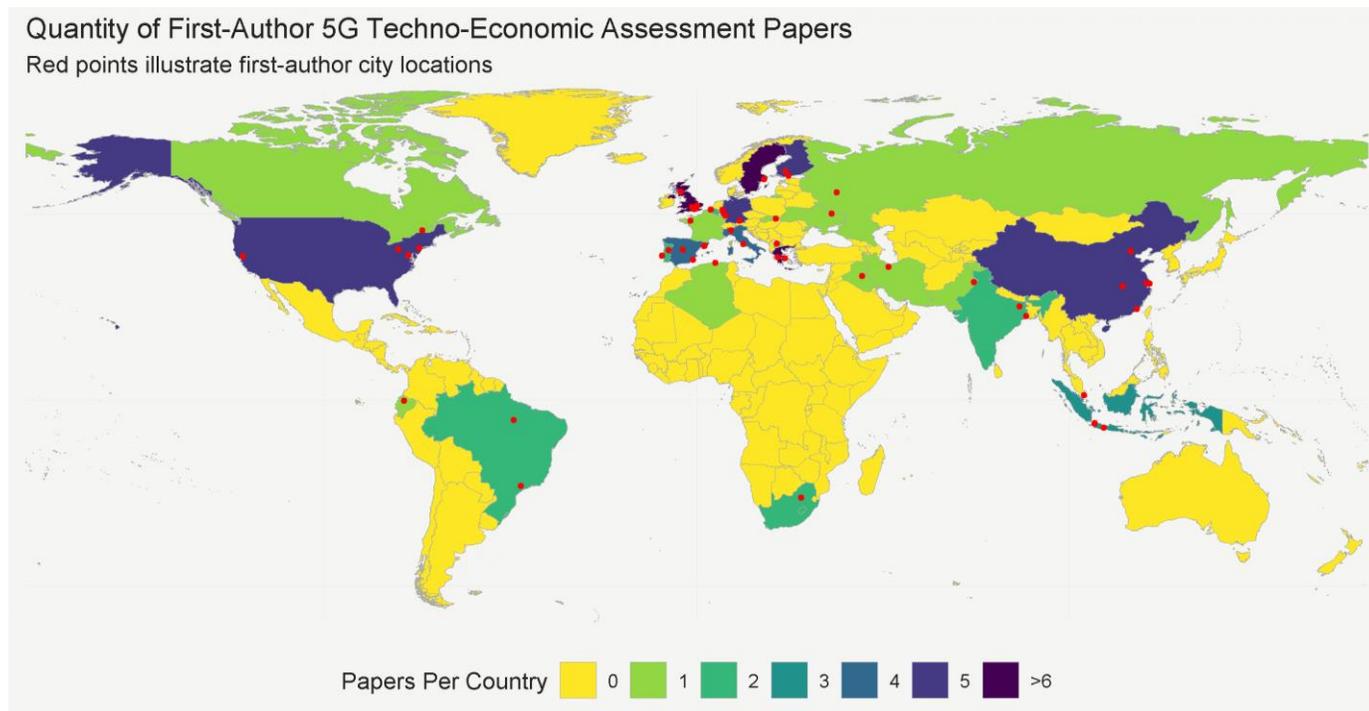

**Fig. 4.** Quantitative summary metrics for the selected 5G TEA sample papers.

For example, existing techno-economic cost comparisons have been undertaken considering the investment requirements for deploying 5G, relative to fixed broadband scenarios (such as Fiber-To-The-Premises) [153]. With 5G providing relatively little improvement in terms of helping to serve rural and remote unconnected citizens, future 6G research needs to be better at identifying cost-efficient technologies prior to standardization to ensure the next generation of cellular technologies can solve global coverage issues.

## V. EVALUATING THE 5G TEA SAMPLE COMPOSITION

Having reviewed the six main areas of the 5G TEA papers, we now present a summary of the composition of the sample, in terms of the publishing outlet and geographic locations of the lead author of each manuscript. This helps to set the overall context for the techno-economic 5G literature, beyond the more technical aspects discussed previously, and is important for drawing recommendations for future research.

Firstly, the number of papers featured by publisher is illustrated in Fig. 3 (F), broken down based on the continent location of the lead author. The IEEE has by far the largest number of publications with more than half the total (approximately 53%), which is unsurprising given it is the premier outlet for communications engineering research. This is then followed by more modest shares across Elsevier (13%), Springer (9%) and Wiley (8%). Specifically, Elsevier contained more techno-economic papers which pertained to policy and economics, mainly due to the large number in Telecommunication Policy, especially following the 5G special issue in 2019. Approximately four paper (5%) exist in IEEE/Optica Publishing Group journals (such as the Journal of Lightwave Technology), focusing mainly on the optical networking aspects of network planning [129], network resilience [156] and backhauling [162].

Secondly, the papers illustrate that the largest quantity of techno-economic research is carried out in Europe (64%), followed by Asia (20%) and the Americas (12%). In contrast, Africa has the second smallest number of papers (4%) included in the sample, whereas no papers were included from Australasia. Particularly the lack of papers produced by African researchers is problematic, because this is one of the continents with the highest number of unconnected broadband users.

The location of the lead-author for each paper has been plotted in Fig. 4, based both on the city location and the overall quantity of 5G TEA papers per country. The most papers were produced in the UK (12%), Greece (9%), Sweden (8%) and Finland (7%). In contrast, the USA and China each produced 5 papers (7% overall each). Surprisingly, Japan or Korea did not feature at all. Many of these trends may be explained by funding priority differences across nations. Indeed, being a leader in the engineering of communications technologies may not necessarily translate to being a leader in the techno-economic analysis of the deployment options. Granted, this analysis is limited by the fact that the sample only includes papers published in the English language.

## VI. DISCUSSION AND RECOMMENDATIONS FOR FUTURE 6G TEA

Having carried out a meta-review of the 5G TEA literature, along the six key themes identified in Fig. 2, we now turn to summarizing high-level lessons. The three research questions articulated earlier in this survey are returned to. As part of this process, the discussion is used in formulating recommendations for the future use of TEA in assessing candidate 6G technologies.

## A. What are the major trends in the use of techno-economic methods for assessing 5G?

A key finding is that most of the 5G techno-economic research carried out has focused on a single use case for Enhanced Mobile Broadband (eMBB), whereas there are relatively few studies which consider either uRLLC or mMTC use cases. Perhaps this reflects the certainty provided by existing 3GPP releases which have largely focused on providing eMBB via the 5G New Radio interface, either using Non-Standalone or Standalone approaches. This could change in the future however, as newer 3GPP releases contain much more of the functionality expected from 5G to enable low latency communications (addressing uRLLC) and industrial IoT (addressing mMTC). This suggests there is still a considerable amount of 5G TEA still to be undertaken, if we are to effectively deliver on the full range of 5G use cases.

Across the four main 5G technology areas identified (network densification, millimeter wave, mMIMO and virtualization), there is a reasonable quantity of research addressing each of these different topics. However, while network densification has been well studied over the past decade for both macro cells and small cells using more traditional architectures, much of the TEA research on virtualization has only been evolving relatively recently. This reflects the emergence of new assessments concentrating on the use of SDN/NFV to enable C-RAN architectures. Often this focuses on high traffic deployment situations, where cost savings could be enabled from virtualization, as demonstrated by the number of papers in the sample focusing exclusively on urban settlement types. This contrasts, with the number of 5G TEA papers considering rural and remote deployment scenarios, which were more limited in number. This lack of focus in the 5G TEA literature could be part of the growing impetus for 6G to pay much greater attention to providing affordable global connectivity and deliver universal broadband [170].

## B. What worked and what did not, in the use of these methods for assessing the techno-economics of 5G?

This review identifies a set of key issues prevalent in the existing 5G literature which will be discussed here. Specifically, four areas have been selected based on their importance to the future development of the TEA research area.

Firstly, the quality of service of an engineered communications system has a very dramatic impact on the TEA results, findings, and take-away messages for 5G research. For examples, assumptions used in the network dimensioning of cellular systems are especially relevant to ascertain the likely accuracy and reliability of the analysis, but those assumptions are often buried in the paper (e.g., not reported in the more obvious places researchers may look when reviewing a paper summary). Future evaluations must be much clearer about these assumptions to other researchers. Ideally, this involves quantifying the sensitivities of any TEA model against heterogenous quality of service assumption sets.

Secondly, the use of meaningful financial metrics varies by paper, with some focusing purely on Capital Expenditure (Capex), such as initial one-off investments in network equipment, and some focusing purely on Operational Expenditure (Opex), such as the energy consumption of different technologies. To understand the cost-efficiency of different 5G technologies, and broader 5G network deployment strategies, researchers need to capture the TCO over a defined deployment period which relates to either asset lifetimes or generational upgrade periods (e.g., cellular generations work mainly on a decadal upgrade basis). This matters because the ongoing shift to 'Anything-As-A-Service' is a key business model development which is pivotal to delivering 5G affordably, but therefore renders examining only one part of the cost equation in isolation limited (e.g., Capex-only analyses).

Another key point identified is that many studies assert that they are focusing on 'cost', although they stop short of presenting potential costs for different strategies. Instead, such assessments opt to only present estimates of asset quantities, which only represent a partial picture of the business implications of different strategic 5G decisions.

Thirdly, there are a range of different model exploration methods which are employed to undertake techno-economic appraisal of 5G networks, including from optimization, sensitivity analysis and scenario analysis. One of the most common techniques used to assess cellular communications networks by engineers is optimization, but we identify that many papers do not always include a sensitivity analysis of the model results presented by the researchers. Since all models are, at best, an approximation of reality that rely on simplifying assumptions, and because the input data may rapidly become dated or is highly context-dependent, presenting sensitivity results is important. This builds on the first point made about embedded quality of service assumptions in 5G TEA. Failure to adequately explore and explain how the modeling results depend on key inputs and modeling decisions (compromises), provides only a partial analysis. Indeed, this leads to significantly less value in using TEA for supporting strategic decisions pertaining to the planning, design, construction, and operation of cellular networks.

Finally, there is generally a poor commitment among researchers working on 5G TEA to share input data, code, and results openly. This is a worrying trend and illustrates this research area is far behind other scientific areas, where submitting open code is a prerequisite to publishing in the most prestigious academic journals (e.g., Science/Nature). Researchers should be confident enough in their analyses to make their data, code, and results openly available to other researchers to enable greater validation of the results, and to lower the barrier to entry for other analysts to comparatively evaluate different 5G techno-economic models.

A key problem with current techno-economic assessment of 5G is that the work is, by definition, multi-disciplinary, covering aspects of engineering and economics. Due to the necessity of addressing two separate areas, researchers must split their time working on multiple domains. Indeed, one of the perils of multi-disciplinary research is that by addressing numerous areas of study simultaneously, the analysis may not stand up to scrutiny by researchers who focus only on a single specialized field of research, given limitations on researcher time. Therefore, it is important that techno-economic research avoids unrealistic modeling simplifications which will not stand up to scrutiny in areas of domain specialization. By this we mean that TEA must satisfy specialists in both engineering *and* economics for any analysis to be credibly received.

*C. How can the use of techno-economic methods be improved for evaluating candidate 6G technologies*

We now reflect on this survey to produce the pivotal contribution of this review. Indeed, this focuses on producing a set of clear recommendations for improving Next-G techno-economic research, particularly regarding 6G R&D, standardization, and prospective deployment. These proposals consist of five key recommendations.

*1) Quality of service assumptions need to be made much clearer to boost research transparency*: The very best 5G TEA papers utilize the actual engineering models used for purely technical analyses. However, when dimensioning a wireless network, the reality is that researchers must make assumptions regarding the spatial and temporal aspects of traffic demand, path loss, interference, spectral efficiency, and various other stochastic factors which influence the level of reliability provided. The challenge of course is that many of these single value assumptions are not necessarily portrayed clearly to the reader, making it hard to understand the potential influence each one may have on the techno-economic results produced.

Therefore, a key recommendation is for future 6G TEA papers to have much clearer quality of service assumptions, for example, pertaining to both traffic demand and the dimensioning of different wireless network architectures. Delivering on this could be more easily achieved if such an endeavor was carried out in tandem with recommendation III (ensuring a sensitivity analysis is provided) and recommendation IV (releasing model inputs, code, and results openly for reproducibility).

*2) Meaningful financial metrics are required to reflect increased network virtualization in both 5G and 6G*: A variety of financial metrics are used to assess the techno-economics of 5G technologies as part of different deployment strategies. However, some more traditional financial approaches do not necessarily reflect the increasing virtualization of the underlying 5G RAN and affiliated transport networks. Therefore, a key recommendation is to eschew capex-only analyses. Indeed, the financial metrics produced by such approaches are not meaningful enough to reflect the increasing use of cloud-processing (via SDN/NFV) in 5G, with this trend set to continue into the 6G paradigm of wireless communications.

*3) Methods utilizing optimization, simulation and scenario analysis must include a sensitivity analysis to quantify 6G model uncertainties*: Some of the most widely used engineering appraisal methods to evaluate the techno-economics of 5G include optimization, simulation, and scenario analysis techniques. This is unsurprising as they form some of the key quantitative methods favored by engineers for supporting decisions pertaining to wireless network design, planning, and operational management. However, a key recommendation from this review is that engineers ensure their assessments of candidate 6G technologies include a sensitivity analysis of any developed models. The purpose of this is to make certain that model uncertainties are fully portrayed to other engineering researchers who have an interest in emerging wireless technologies, to avoid misinterpretation of key 6G performance metrics, especially for different deployment contexts.

*4) Researchers must be confident enough in their analysis to openly share 6G model data and code*: In other areas of science, there is a significant effort to ensure published work in the most prestigious journals provides all model input data, code, and results for other researchers to evaluate, validate and scrutinize key contributions. This is to ensure the highest standards in scientific enquiry, as well as proper attribution, should errors be present (either unintentionally or intentionally). Transparency of this nature is all part of the scientific process of enquiry and dissemination and should not be shied away from. Indeed, researchers not willing and open to taking these steps when publishing in the most prestigious engineering journals should raise concerns. We recommend the IEEE and other key engineering outlets adopt a similar open science policy to other key scientific publications such as Science and Nature requiring researchers to make their analyses openly available to all. This would ensure science contributes as effectively as possible to societal progress, particularly as a considerable amount of engineering research is publicly funded via taxation from citizens (and therefore should be freely available for all).

*5) Greater multi-disciplinary collaboration is required during the R&D and standardization process for ongoing techno-economic assessment of candidate 6G wireless technologies*: All multi-disciplinary research can face collaboration challenges, and the evaluation of 5G and 6G techno-economic research is no different. Our conjecture is that 6G requires greater focus on techno-economic appraisal as part of the early R&D and standardization process, not just in the deployment phase, as we have seen for 5G. Ultimately, this is too late to help to change the proposed range of technological options. Thus, we demonstrate the importance of this recommendation by reflecting on the affordability of 5G technologies, and what it means for 6G.

Deploying broadband connectivity in rural and remote locations is already emerging as an important agenda goal for 6G. This is reflective of the fact that 5G places less of a focus on solving rural coverage issues, and greater focus on increasing the capacity of cellular networks in urban and suburban locations. We agree that serving urban deployment scenarios is crucial to meeting growing traffic demand, although it does provide a major strategic shortcoming of this generation of cellular technologies. Indeed, some have been arguing for many years that it would be more desirable for 5G to provide consistent and reliable coverage, not ever faster per user capacities [171]. This is not necessarily an either-or prospect, and therefore 6G could quite easily have both capacity-enhancing and coverage-enhancing streams of research.

With this debate gaining traction in 6G, the affordability of new cellular systems is a common theme in Next-G position papers. A key use case is attempting to viably provide broadband services to the other ~50% of the global population who are yet to connect to the Internet (or are currently under-served by existing infrastructure). Therefore, it is imperative that engineers focus more on considering the affordability implications of candidate 6G technologies, at least at a stage prior to 6G standardization, particularly for rural broadband use cases. Such an approach can be central to the 6G research which aims to develop lower cost cellular technologies for deployment in ARPU-constrained locations. Thus, the effective TEA of candidate 6G technologies must take place simultaneously with technical engineering and computer science research to ensure such affordability metrics can be achieved.

To overcome this issue, we recommend greater collaboration between the leading engineers and economists working on Next-G technologies, such as 6G. Indeed, research funding institutions rightfully recognize the importance of fundamental engineering research. Yet, they should also consider allocating small amounts of funding for multi-disciplinary research which includes quantitative analysis of the cost ramifications of new communications innovations. Such an activity can generate insight which can help to direct future fundamental R&D pursuits.

## VII. CONCLUSIONS

This survey has evaluated the techno-economic 5G literature based on six key dimensions. These cover 5G use cases, technologies, modeling techniques, financial metrics, business models, and the spatial focus of each assessment. In total, 150 papers were identified for potential evaluation, with 75 being selected for the final sample constituting this detailed meta-review. Importantly, the analysis climaxed with five key recommendations pertaining to the future techno-economic appraisal of candidate 6G wireless technologies. These proposals reflect modeling uncertainties, metric usefulness, the increasing use of virtualization, the open science agenda, and a call for greater multi-disciplinary collaboration as part of the 6G R&D and standardization process.

To finalize, we identify a key limitation of this analysis. The assessment was confined to only those papers published in the English language which may mean that the sample is underrepresented by some contributing countries. Therefore, future analyses of the techno-economic literature may benefit from having co-authors who can also review publications in Mandarin, Korean and Japanese.

## ACKNOWLEDGMENT

The authors would like to thank attendees of TPRC49 for constructive feedback on an early version of the paper, which helped to shape the structure of the final piece. Additionally, the authors thank the multiple anonymous referees for reviewing the manuscript.

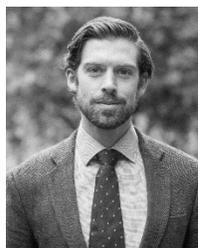

**Edward J. Oughton** received the M.Phil. and Ph.D. degrees from Clare College, at the University of Cambridge, U.K., in 2010 and 2015, respectively. He later held research positions at both Cambridge and Oxford. He is currently an Assistant Professor in the College of Science at George Mason University, Fairfax, VA, USA, developing open-source research software to analyze digital infrastructure deployment strategies. He received the Pacific Telecommunication Council Young Scholars Award in 2019, Best Paper Award 2019 from the Society of Risk Analysis, and the TPRC48 Charles Benton Early Career Scholar Award 2021.

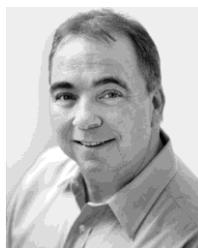

William Lehr (Senior Member, IEEE) received the B.A., B.S., and M.S.E. degrees from the University of Pennsylvania, the M.B.A. degree in finance from the Wharton School, and the Ph.D. degree in economics from Stanford University. He is currently an Internet and Telecommunications Industry Economist and a Research Scientist with the Computer Science and Artificial Intelligence Laboratory (CSAIL), Massachusetts Institute of Technology (MIT), where he is also a part of the Advanced Network Architectures Group. He advises public and private sector clients in U.S., and abroad on ICT strategy and policy matters. His research interests include economic implications of ICT technologies for public policy, industry structure, and evolving internet ecosystems.